\documentclass[prl,aps,showpacs,lengthcheck,superscriptaddress,twocolumn]{revtex4}
\usepackage{amsmath}
\pdfoutput=1
\usepackage{xcolor}
\usepackage{amsfonts}
\usepackage{graphicx}
\usepackage{bm}

\newcommand{\black}[1]{{\color{black} #1}}

\newcommand{\beq}{\begin{equation}}
\newcommand{\eeq}{\end{equation}}
\newcommand{\beqa}{\begin{eqnarray}}
\newcommand{\eeqa}{\end{eqnarray}}

\begin{document}

\title{Hydrodynamics of Turning Flocks}

\author{Xingbo Yang}
\affiliation{Physics Department, Syracuse University, Syracuse NY 13244, USA}
\author{M. Cristina Marchetti}
\affiliation{Physics Department, Syracuse University, Syracuse NY 13244, USA}
\affiliation{Syracuse Biomaterials Institute, Syracuse University, Syracuse NY 13244, USA}

\begin{abstract}
{We present a hydrodynamic model of flocking that generalizes the familiar Toner-Tu equations to incorporate turning inertia of well-polarized flocks. The continuum equations controlled by only two dimensionless parameters, orientational inertia and alignment strength, are derived by coarse graining the inertial spin model recently proposed by Cavagna et al.~\cite{Cavagna2014}. The interplay between orientational inertia and bend elasticity of the flock yields anisotropic spin waves that mediate the propagation of turning information throughout the flock. The coupling between spin current density to the local vorticity field through a nonlinear friction gives rise to a hydrodynamic mode with angular-dependent propagation speed at long wavelength. This mode goes unstable as a result of the growth of bend and splay deformations augmented by the spin wave, signaling the transition to complex spatio-temporal patterns of continuously turning and \black{swirling} flocks. }

\end{abstract}

\maketitle

The Vicsek model~\cite{Reynolds1987,Vicsek1995} and related continuous-time variations~\cite{Romanczuk2012} have been used to model flocking in a variety of systems, from birds~\cite{Ballerini2008} to cells~\cite{Szabo2006} to \emph{in vitro} cellular components~\cite{Schaller2010,Sumino2012} and synthetic swimmers~\cite{Bartolo2013}.
These are examples of active systems, consisting of  individually driven, dissipative units that exhibit coordinated motion (flocking) at large scales~\cite{Marchetti2013,Ramaswamy2010}. In the Vicsek model the active units are described as point particles with overdamped dynamics carrying a velocity vector of fixed magnitude, hence ``flying spins''. Each spin tends to align with  its neighbors, but makes errors, modeled as angular noise~\cite{Vicsek1995}. 
{\color{black} The system exhibits a liquid-gas phase transition from a disordered gas state to a polar liquid state as the noise is decreased or the number density is increased, with microphase separation in the coexistence region \cite{Solon2015}.} The existence of the transition has been put on firm grounds by a large number of numerical studies~\cite{Vicsek1995,Gregoire2004}.  Toner and Tu also proposed a continuum version of the model inspired by dynamical field theories of condensed matter systems~\cite{Toner1995,Toner2005}. 

Recent work \cite{Attanasi2014} has suggested that the description of the observed collective turning of bird flocks requires a modification of the Vicsek model to include angular inertia in the dynamics. This allows propagation of angular correlations through the flock on large scales via spin-wave-like excitations~\cite{Cavagna2014}. 
In this paper we derive the continuum equations for such an  ``inertial spin model" by explicitly coarse-graining the microscopic dynamics. The resulting equations (Eqs.~\ref{eq:rho}-\ref{eq:spin}) generalize the Toner-Tu model to account for turning modes by incorporating the dynamics of the spin angular momentum of the flock. These equations, governed by only \black{two dimensionless parameters}, are the first important result of our work. They contain new terms as compared to the phenomenological model of Ref.~\cite{Andrea2015}, most importantly a nonlinear friction that couples spin and density fluctuations to bend and splay deformations of the order parameter.
{\color{black} This new coupling transforms the propagating density bands ubiquitously observed in flocking models into turning bands of spin currents, driving the transition to a novel state of  continuously swirling and rotating flocks, where turning information are transmitted by anisotropic propagating spin waves. The predicted sound speeds could in principle be measured in experiments.}


%
\begin{figure}[!h]
\includegraphics[width=1.00\columnwidth]{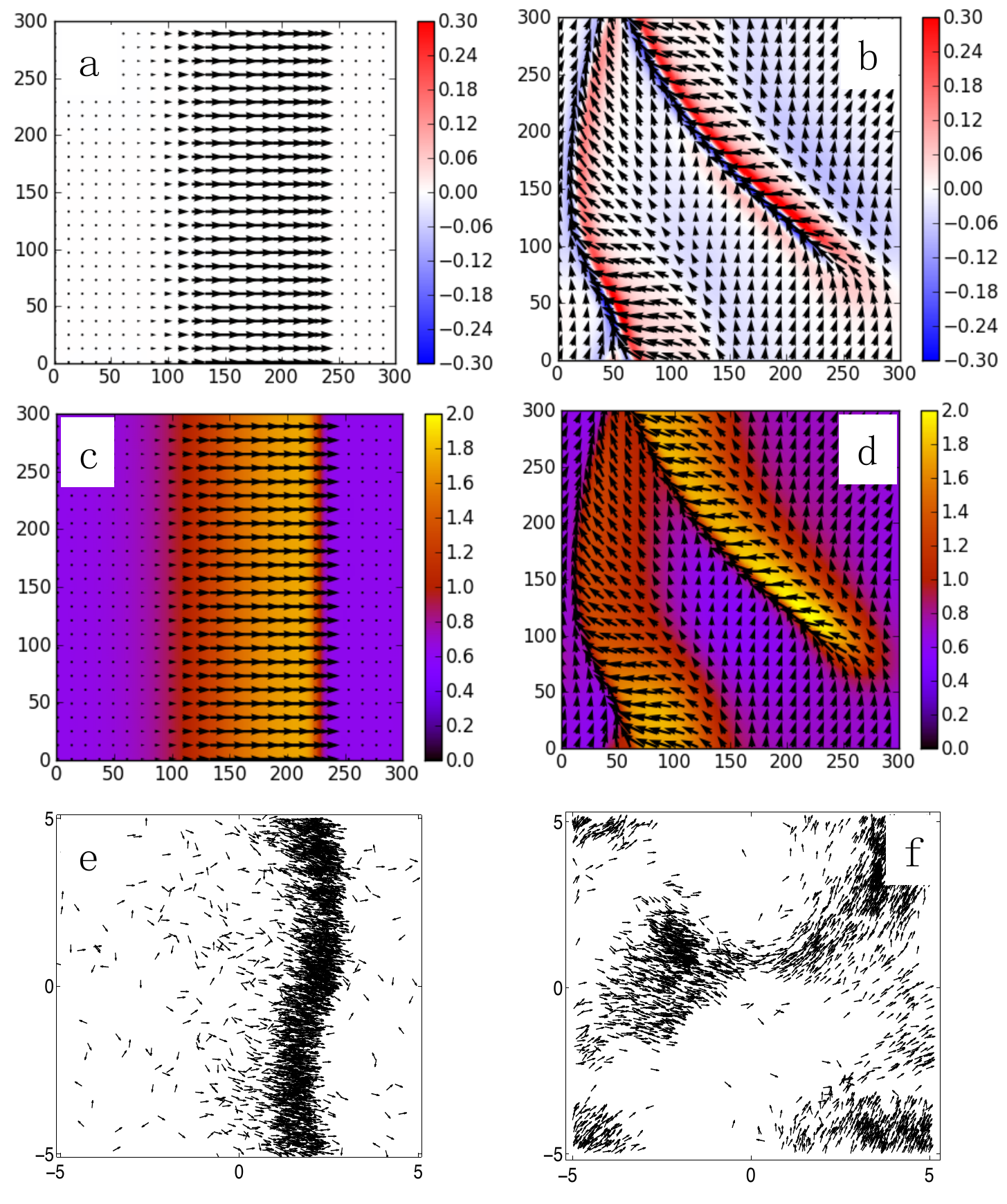}
\caption {Flocking patterns obtained via numerical solutions of Eqs.~\ref{eq:rho} -\ref{eq:spin} \black{(top \& middle rows) and via} particle simulations using Eqs.\ref{eq:ri}-\ref{eq:si} (bottom row). The \black{left/right} columns show the patterns obtained at points $A/B$ of the phase diagram Fig.\ref{fig:PD}a, corresponding to  {\color{black} the spinodal portion of the coexistence region of traveling bands of polar liquid and disordered gas and to the region of the spin-wave instability}, respectively. \black{The snapshots (a-d) are obtained with  $\tilde{\gamma}=2.0$, $\tilde{\chi}=0.5$ (a,c) and $\tilde{\gamma}=7.6$, $\tilde{\chi}=1.7$ (b,d) on a 300 by 300 grid lattice with grid size $0.1$, integration time step $0.002$ and periodic boundary conditions. The particle simulations are performed with $3000$ particles in a box of size $L=10$ with periodic boundary conditions. Simulation parameters are $R=1$, $\epsilon=2.0$, $v_0=2.0$, $\chi=1.0$, $\eta=1.0$ and $\gamma=0.16$ (e) and $\gamma=0.80$ (f). The integration time step is $0.01$.} a-d: the arrows represent the local polarization, with length proportional to the polarization strength. The color indicates spin current density in a-b and number density in c-d.  e-f: the arrows represent polarization of individual particles (see Supplementary Movies).}
\label{fig:numerical}
\end{figure}

Our starting point is the continuous-time  model of inertial spins proposed by Cavagna \emph{et al.}~\cite{Cavagna2014}, where $N$ point particles in a two-dimensional box of area $L^2$, with average number density $\rho_0=N/L^2$,  interact via a pairwise aligning interaction.
Each particle  is described by its  position $\bm r_i$ and the direction of its velocity, identified by  a unit vector $\hat{\bm e}_{\theta_i}=\left(\cos\theta_i,\sin\theta_i\right)$ in 2D.  The dynamics of the  $i$-th spin is described by
\begin{gather}
\label{eq:ri}
\frac{d\bm r_i}{dt}=v_0\bm \hat{\bm e}_{\theta_i}\;,~~~~~\frac{d\theta_i}{dt}=\frac{1}{\chi}s_i\;,\\
\label{eq:si}
\frac{ds_i}{dt}=\gamma\sum\limits_{j}\tilde{F}(\theta_j-\theta_i,\bm r_{ji})-\frac{\eta}{\chi}s_i+\sqrt{2\epsilon}~ \xi_i(t)\;,
\end{gather}
with $\bm r_{ji}=\bm r_j-\bm r_i$, $v_0$ the self-propulsion speed, $s_i$ the spin angular momentum and $\chi$ the spin moment of inertia.  The spin is an internal angular momentum that generates the self-rotation, and is distinct from the angular momentum of the center of mass.  The polar aligning interaction of strength $\gamma$ is given by $\tilde{F}(\theta,\bm r)=\sin(\theta)/(\pi R^2)$ if $|\bm r|\le R$ and zero otherwise, with $R$ the range of interaction.  This form of the interaction used before in the literature~\cite{Farrell2012} allows us to make analytical progress in the derivation of the continuum equations.
Finally, $\eta$ is a friction and $\epsilon$ describes the strength of the angular noise, with $ \xi_i(t)$ a Gaussian white noise with zero mean and unit variance. 

On time scales large compared to the relaxation time $\tau_\eta=\chi/\eta$, one can neglect the time derivative on the left hand side of Eq.~\eqref{eq:si} and eliminate the spin angular momentum, $s_i$, from the angular dynamics. This yields a continuous-time version of the Vicsek model, with effective alignment strength $\gamma/\eta$ and effective angular noise $\epsilon/\eta^2$.
Two additional time scales govern the dynamics of the system:  the effective rotational diffusion time, $\tau_{\epsilon}=\eta^2/\epsilon$, and the alignment time, $\tau_{\gamma}=\eta/(\gamma\rho_0)$.



Following standard methods~\cite{Dean1996,Zwanzig2001}, one obtains the noise-averaged Fokker-Planck equation associated with the microscopic dynamics described by Eqs.~\eqref{eq:ri} and \eqref{eq:si}, as
\begin{gather}
\label{eq:FP}
\left(\mathcal{D}_t+\frac{s}{\chi}\partial_\theta\right){P}=\partial_s\left[(\eta\frac{s}{\chi}+T[P])P\right]+\epsilon\partial^2_s P\;,
\end{gather}
where $\mathcal{D}_t=\partial_t+v_0\bm e_{\theta}\cdot\bm \nabla$ is the material derivative, $P(\bm r,\theta,s,t)$ is the  probability density of particles at position $\bm r$, with velocity in direction $\theta$ and spin $s$ at time $t$, and $T[P]$ is the aligning torque
\begin{gather}
\label{eq:tau}
T[P]=-\gamma\int_{\theta'}\int_{s'}F(\theta'-\theta)P(\bm r,\theta',s',t)\;.
\end{gather}
For simplicity we have assumed $\tilde{F}(\theta,\bm r)=\delta(\bm r)F(\theta)$, with $F(\theta)=\sin(\theta)$, \black{neglecting interaction between pairs at different positions}.

We describe the large-scale dynamics in terms of a few coarse-grained fields that vary slowly relative to microscopic time scales. For polarized flocks in addition to the number density, $\rho(\bm r,t)$, of active units and their polarization current density, $\bm w(\bm r,t)$, we include the spin angular momentum density, $S(\bm r, t)$. These are obtained from the probability density $P$ as
\begin{gather}
\label{eq:fields}
\left(\begin{array}{c}
\rho(\bm r,t)\\
\bm w(\bm r,t)\\
S(\bm r,t)
\end{array} \right)
=\int_\theta\int_s 
\left(\begin{array}{c}
1\\
\bm{\hat{e}}_\theta \\
s
\end{array}\right)
P(\bm r,\theta, s,t)\;.
\end{gather}
To obtain a closed set of hydrodynamic equations for $\rho$, $\bm w$ and $\bm S= S\bm{\hat z}$, we combine moment techniques used to approximate the velocity-dependent part of the Fokker-Planck equation~\cite{Risken1988} with the closure developed in Ref.~\onlinecite{Bertin2009,Chate2014} to handle kinetic equations of active systems (see Supplementary Material). To minimize the number of parameters, we nondimensionalize the equations by scaling time with $\tau_{\epsilon}=\eta^2/\epsilon$, length with \black{$v_0\tau_{\epsilon}$} and density with $\rho_0=N/L^2$. The resulting equations are controlled by only two dimensionless parameters $\tilde{\chi}=\tau_{\eta}/\tau_{\epsilon}$ and $\tilde{\gamma}=\tau_{\epsilon}/\tau_{\gamma}$\black{~\footnote{See the SI for a description of our choice of parameters.}}. For simplicity, we drop the tildes and all parameters are dimensionless in the following discussion unless otherwise noted. The continuum equations are given by
\begin{gather}
\label{eq:rho}
\frac{\partial \rho}{\partial t}=-\bm \nabla\cdot\bm w\;,\\
\label{eq:w}
\mathcal{D}_t^{w}\bm w=-\left[\alpha(\rho)+\beta|\bm w|^2\right]\bm w-\frac{1}{2}\bm \nabla\rho+\lambda_2\bm w(\bm \nabla\cdot\bm w)\notag\\+\Omega_1\bm S\times\bm w+\Omega_2\bm \nabla\times\bm S+D_w\nabla^2\bm w\;,\\ 
\mathcal{D}_t^{s}\bm S=-\bm \nabla\times\left[\left(\alpha(\rho)+\beta|w|^2\right)\bm w\right]+\Omega_3\bm w\times\nabla^2\bm w\notag\\-\lambda_s(\bm \nabla\cdot\bm w)\bm S-\xi\bm S+D_s\nabla^2\bm S\;,
\label{eq:spin}
\end{gather}
%
where $\mathcal{D}_t^w=\partial_t+\lambda_1\bm w\cdot\bm \nabla$ and $\mathcal{D}_t^s=\partial_t+\lambda_s\bm w\cdot\bm\nabla$ are convective derivatives, $\alpha(\rho)=(1-\frac{\gamma\rho}{2})/(1+\chi)$, $\beta=\gamma^2/[8(1+\chi)]$ \black{and $\xi=1/\chi$.} Explicit expressions for all other dimensionless parameters are given in the supplementary material. A pressure-type term $\lambda_3\nabla|w|^2$ has been neglected in Eq.(7) because this term is known to lead to a spurious instability even in the overdamped limit when $\lambda_3$ is evaluated with the closure used here \cite{Mishra2010,Chate2014}. This instability has not been observed in particle simulations of Vicsek models. We have also verified that it is not obtained in particle simulations of the inertial spin model.

Equations (\ref{eq:rho}-\ref{eq:spin}) augment the flocking model of Toner and Tu~\cite{Toner1995} by incorporating the dynamics of the spin current. When $\bm S$ is neglected, these equations reduce to the Toner-Tu equations as derived  by Farrell \emph{et al.}~\cite{Farrell2012} (but in the case of constant self-propulsion speed). As in the Toner-Tu model, the vector field  $\bm w$ plays the dual role of polarization density and flow velocity. In equilibrium systems of rotors both the equations for the spin and the velocity field $v_0\bm w$ would contain dissipative couplings describing friction with the substrate proportional to the combination $\bm{S}/\chi-\frac{v_0}{2}\bm\nabla\times\bm w$, guaranteeing that the angular velocity $\bm S/\chi$ and the vorticity $\frac{v_0}{2}\bm\nabla\times\bm w$  be equal when the whole system is rotating as a rigid body~\cite{Lubensky2005,Braun1997}. In the nonequilibrium system considered here, in contrast, frictional terms proportional to angular velocity and vorticity will in general appear with different coefficients. The first term on the right hand side of Eq.~\eqref{eq:spin} was not included in previous phenomenological model \cite{Andrea2015} and has  a natural interpretation of a nonlinear, velocity-dependent vortical friction.
The  ``self-spinning''  term $\bm S\times\bm w$ couples the center-of-mass motion to the turning dynamics. In contrast to systems of passive rotors~\cite{Weinberg1977,Lubensky2005}, in the self-propelled  particle model considered here, these two degrees of freedom are coupled because the spinning angle also controls the direction of translational motion~\cite{Cavagna2014}. 
We expect these equations will provide useful to describe a number of active systems where collective turning controls the large-scale dynamics.

The homogeneous steady states of the continuum equations have uniform density, $\rho=1$, and zero mean value of the spin, $\bm S=0$. As in the Toner-Tu model with no angular inertia, there are two such states:  an {\color{black}isotropic gas state},  with $\bm w=0$, and a {\color{black} polarized liquid} or flocking state, with  $\bm w=w_0\bm{\hat{x}}$ and  $w_0=\sqrt{-\alpha_0/\beta}$, where $\alpha_0=\alpha(\rho=1)$. We have chosen the $\bm{\hat{x}}$ axis along the direction of spontaneously broken symmetry. The isotropic state is always \black{linearly stable for $\gamma<2$.}
We examine below the linear stability of the polarized state by considering the dynamics of fluctuations. We let $\bm w=w_0 \hat{\bm x}+\delta\bm w$, $\rho=1+\delta \rho$, $\bm S=\hat{\bm z}\delta s$ and introduce Fourier amplitudes $(\delta \rho,\delta \bm w,\delta s)=\sum_{\bm q}(\rho_{\bm q},\bm w_{\bm q},s_{\bm q})e^{i\bm q\cdot \bm r+\sigma t}$ to obtain a set of linearized equations in Fourier space (see Supplementary Material).

\begin{figure}[!h]
\includegraphics[width=1.00\columnwidth]{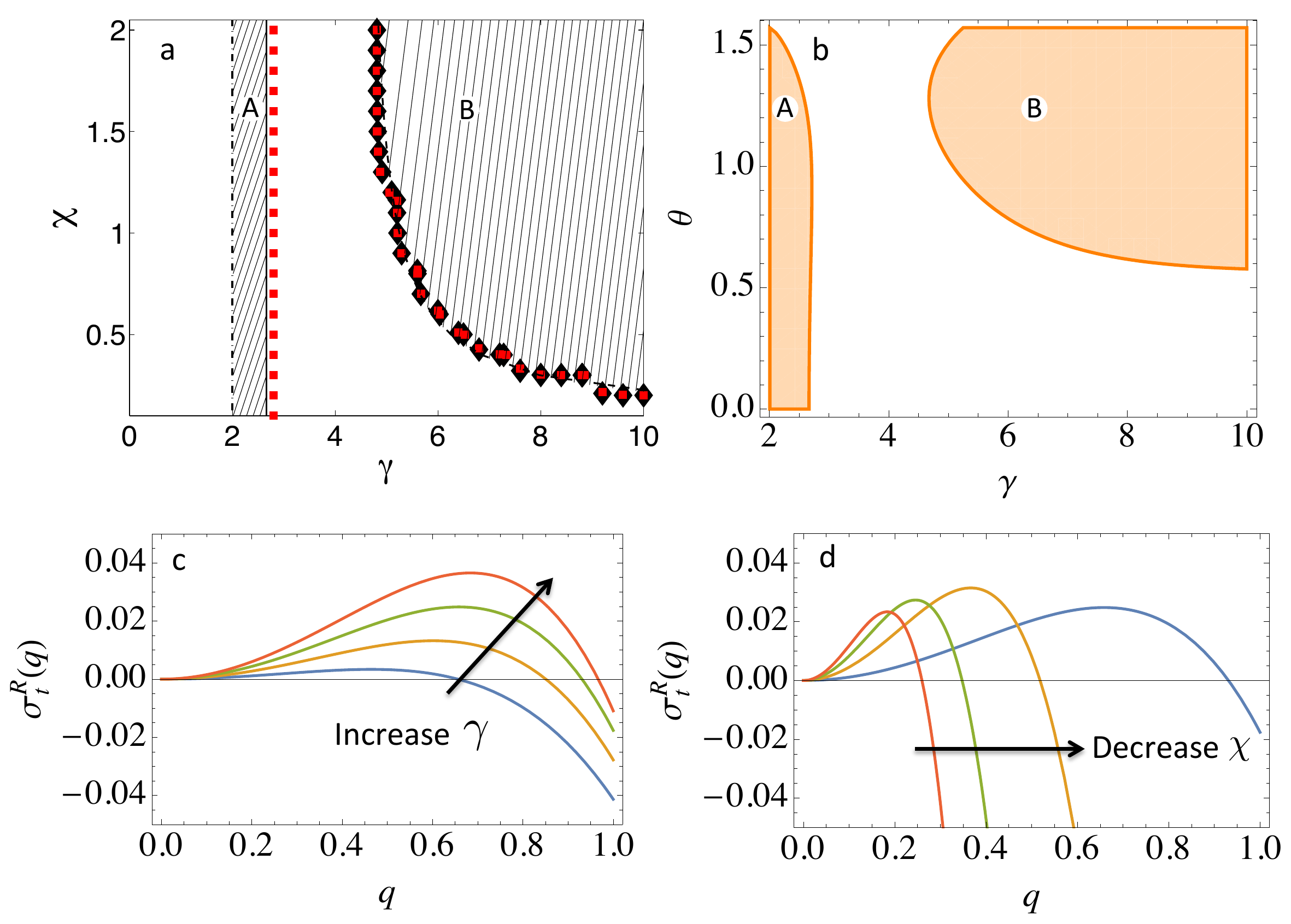}
\caption{a. Phase diagram in the plane of dimensionless $\gamma$ and $\chi$. b. Phase diagram in the plane of dimensionless $\gamma$ and $\theta$ at $\chi=1.0$. {\color{black}  The shaded region A is the spinodal portion of the region of  coexistence of disordered gas (existing for $\gamma<2$) and traveling bands of polar liquid (existing for $\gamma>8/3$). The coexistence region is delimited by the binodals (not shown) and extends inside the white regions, both to the right and to the left of region A, as verified via particle simulations. The shaded region  B corresponds to the region where the homogeneous polar liquid is linearly unstable to spin-waves} as shown in Fig.\ref{fig:numerical}. c. Real part of the dispersion relation of the transverse mode $\sigma_t^{\pm}=\sigma^{\pm}(\pi/2)$ at $\chi=1$ and $\gamma=7.0, 8.0, 9.0, 10.0$. d. Real part of the dispersion relation of the transverse mode $\sigma_t^{\pm}=\sigma^{\pm}(\pi/2)$ at $\gamma=9$ and $\chi=0.5, 1.0, 1.5, 2.0$.}
\label{fig:PD}
\end{figure}

For spatial variations along the direction of broken symmetry ($\bm q=q\bm\hat{\bm x}$),
$w_{\bm q}^y$ and $s_{\bm q}$ decouple from $\rho_{\bm q}$ and $w_{\bm q}^x$. 
The coupled linear dynamics of fluctuations in the density and the magnitude of polarization ($w_{\bm q}^x$) is unaffected by angular inertia and is controlled by a longitudinal propagating mode, with propagation speed $c_\rho=|\alpha_{\rho}|/(2\beta w_0)$, where $\alpha_{\rho}=\partial_{\rho}\alpha$. This mode goes unstable when $\gamma<8/3$, corresponding to region A in Fig.\ref{fig:PD}a. This instability is known in Vicsek and Toner-Tu models as banding instability, {\color{black} but has recently been identified as the spinodal boundary within the liquid-gas coexistence region}  (Fig.1 left column) ~\cite{Bertin2006,Bertin2009,Mishra2010,Solon2015}. 
The coupled dynamics of spin and bending fluctuations ($w_{\bm q}^y$) gives rise to overdamped, finite-wavelength spin waves that mediate the propagation of turning information throughout the flock with wave speed $c_s=w_0\sqrt{\Omega_1\Omega_3}$ that increases with alignment strength. The existence of such propagating spin waves  has been demonstrated on the basis of general arguments~\cite{Cavagna2014} and phenomenological continuum models~\cite{Andrea2015}, where they were dubbed ``second sound''. 

For wavevectors along any directions other than the direction of broken symmetry, all four equations are coupled and the analysis of the modes is rather cumbersome. For small wavevectors, we find two
stable and relaxational modes that will not be discussed further and two hydrodynamic propagating modes, with dispersion relation
%
\begin{gather}
\label{eq:spin-t}
\sigma^{\pm}(q,\theta)=\ i c^\pm(\theta)q-\mathcal{D}_{sw}(\theta)q^2+\mathcal{O}(q^3)\;,
\end{gather}
and wave velocity
\begin{gather}
\label{eq:cpm}
c^{\pm}(\theta) = \frac{\alpha_{\rho } \cos (\theta )\pm \sqrt{\alpha _{\rho }^2 \cos ^2(\theta )+8 \beta ^2 w_0^2 \sin ^2(\theta )}}{4 \beta w_0}, 
\end{gather}
where $\theta$ is the angle between the direction of $\mathbf{q}$ and the direction of broken symmetry. The full expression for the damping $\mathcal{D}_{sw}(\theta)$ is not instructive thus is not given here.
For $\theta=0$, $c^-(0)=\alpha_{\rho}/(2\beta w_0)$ and the mode $\sigma^-$ yields the banding instability {\color{black} that delimits the spinodal region of microphase separation~\cite{Solon2015}}. For arbitrary angle $\theta$, however, both modes are propagating with anisotropic speed and describe coupled fluctuations of density, spin, and bend/splay deformations of the polarization field. The angular dependence of the instability is shown in Fig.\ref{fig:PD}b that displays the regions where $\mathcal{D}_{sw}<0$. At small angles the instability is driven by density fluctuations, as in the Toner-Tu model. At large angles the instability is dominated by spin fluctuations. At $\theta=\pi/2$ the longitudinal banding instability is suppressed and the dynamics is controlled by transverse spin wave propagating at speed $|c^\pm(\pi/2)|=1/\sqrt{2}$. 
In terms of our dimensionless parameters, this transverse spin wave is unstable for $\gamma>(1+4\chi)(1+\chi)/(8\chi^2)+4$, corresponding to region B in Fig.\ref{fig:PD}a. The instability is driven by the growth of bend $-\bm \nabla\times\left[\left(\alpha(\rho)+\beta|w|^2\right)\bm w\right]$ and splay $\lambda_2\bm w(\bm\nabla\cdot\bm w)$ deformations augmented by the spin wave through the self-rotation term $\Omega_1 \bm S\times \bm w$. This long-wavelength instability of the ordered state is a new result of our work and  will be referred to as spin-wave instability. It leads to a complex spatio-temporal dynamics with large density and spin fluctuations characterized by continuously turning and swirling flocks as confirmed by numerical solutions of the hydrodynamic equations and particle simulations (see Fig.\ref{fig:numerical} right column). 
 
By carrying out the small wavevector expansion of the dispersion relation Eq~\eqref{eq:spin-t}  up to fourth order in  $q$ we can identify the wavector $q_c$ of the fastest growing mode corresponding to the maximum of $Re[\sigma_t^{\pm}(q)]$ shown in  Fig.~\ref{fig:PD}c and d for various values of $\gamma$ and $\chi$. \black{This defines the characteristic length scale $\lambda_c \sim 1/q_c$ that can be thought of as controlling the size of the turning flock {\color{black} at the linear level}.}
\begin{figure}[!h]
\includegraphics[width=1.00\columnwidth]{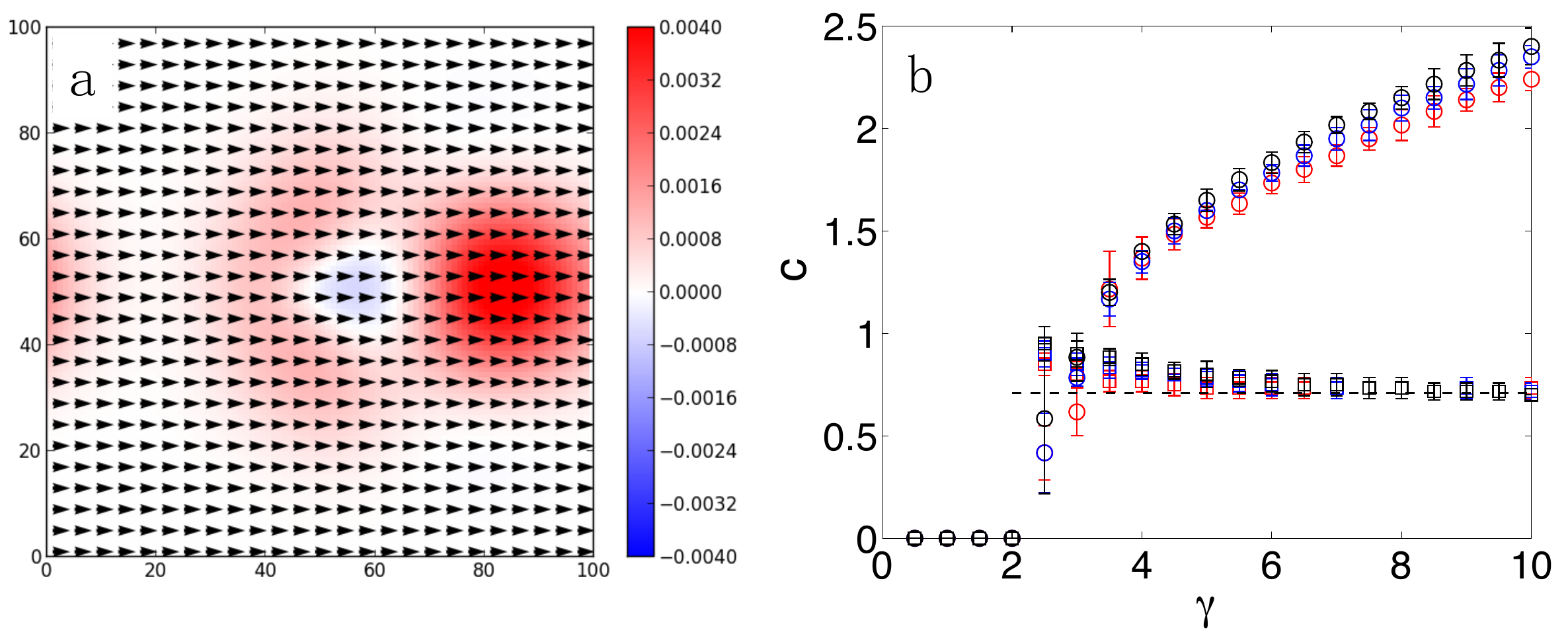}
\caption{a: Snapshot of the anisotropic spin wave in the polarized state at $\gamma=7.0$ and $\chi=2.0$. Color indicates the spin current density. b: Speed of spin waves in the polarized state as a function of alignment strength $\gamma$ for $\chi=1.0,1.5,2.0$ (red, blue, black) in the directions longitudinal (circles) and transverse (squares) to that of mean polarization. The dashed line is the transverse speed $|c^{\pm}(\pi/2)|$ in Eqn.\ref{eq:cpm}. The system is evolved for 5000 time steps.}
\label{fig:wave_speed}
\end{figure}

To gain more insight on the complex spatio-temporal structures that emerge in the unstable regions of parameters and to confirm the results of the linear stability analysis, we have solved numerically Eqs.~(\ref{eq:rho}-\ref{eq:spin}) with periodic boundary conditions {\color{black}starting from the homogeneous polar state with small perturbations}. 
The results are summarized in the phase diagram of Fig.\ref{fig:PD}a. {\color{black} The shaded region A is bounded to the left by the line $\gamma=2$ where the disordered gas is linearly unstable and to the right by the line $\gamma=8/3$ where the homogeneous polar liquid is linearly unstable to longitudinal fluctuations (the banding instability). These instability lines delimit the spinodal portion of the gas/liquid coexistence region and are distinct from the binodal lines that mark the boundaries of such a region~\cite{Solon2015}. In fact particle simulations reveal that the coexistence region extends to the left and right of region A. The squares in Fig.\ref{fig:PD}a correspond to mean density fluctuations  $\Delta {\rho}=\sqrt{\frac{1}{N}\sum_{\bf r}<(\rho(\bm r)-\rho_0)^2>}/\rho_0=0.003$, with $N$ the number of grid points, evaluated in the continuum model, starting in a uniform polar state. The shaded region B  is the region where the homogenous polar liquid is linearly unstable to spin wave fluctuations. 
Again, particle simulations show that the inhomogeneous spinning bands are found beyond the linear stability boundary that delimits region B, suggesting that this region is also a spinodal region.} {\color{black}The diamonds correspond to  spin fluctuations $\Delta S=\sqrt{\frac{1}{N}\sum_{\bf r}<(S(\bm r)-<S>)^2>}=0.0003$}. \black{In the overdamped limit $\chi\rightarrow 0$, the spin-wave instability vanishes due to the rapid decay of spin current fluctuations over time $\chi/\eta$, and the dynamics of the system is controlled solely by a rescaled alignment strength $\gamma$, with a generic banding instability close to the flocking transition, as in the Vicsek and the Toner-Tu models ~\cite{Bertin2006,Bertin2009,Mishra2010}. Our result, together with Ref.~\cite{Andrea2015}, highlights for the first time the importance of inertia in controlling dynamics of active polar systems at large length scales.}

To understand the nature of the spin waves that mediate the transfer of turning information within the flock, we study the propagation of the spin waves numerically with Eqs.~(\ref{eq:rho}-\ref{eq:spin}) by initializing the system in the uniformly polarized state, with a concentrated spin current at the center (Fig.~\ref{fig:wave_speed}a). We measure the longitudinal and transverse speed as a function of alignment strength $\gamma$ for various $\chi$ and plot the results in Fig.~\ref{fig:wave_speed}b. The longitudinal speed (circles) increases with the strength of alignment interaction while the transverse speed (squares) stays approximately constant over the range of parameters.

In the longitudinal direction, where $\delta w_y$ and $\delta s$ decouple from $\delta w_x$ and $\delta \rho$, the spin wave is governed by a damped wave equation at finite wavelength with wave speed $c_s=w_0\sqrt{\Omega_1\Omega_3}$ proportional to alignment strength. In the transverse direction, all fluctuations are coupled and the dynamics is governed at long wavelength by the hydrodynamic mode (Eq.\ref{eq:spin-t})
with an angular-dependent propagating speed that reduces to $|c^{\pm}(\pi/2)|=1/\sqrt{2}$ in the transverse direction as given in Eqn.\ref{eq:cpm}, and fits the data quantitatively in Fig.\ref{fig:wave_speed}b. 


We have derived continuum equations that generalize the Toner-Tu model of flocking to incorporate turning inertia by coarse-graining the active inertial spin model proposed recently by Cavagna \emph{et al. }\cite{Cavagna2014}.  The coarse-graining simplifies the analysis by shrinking the number of independent parameters to two. The interplay between rotational inertia and bending elasticity of a polarized flock provides a mechanism for the propagation of turning information through the flock in the form of collective spin-wave excitations. By studying the continuum equations analytically and numerically, we predict a new instability of the polarized state associated with large density and spin current fluctuations that leads to complex spatio-temporal patterns of continuously swirling and rotating flocks. This long-wavelength instability is associated with the growth of anisotropic spin waves and is referred to as\emph{ spin-wave instability}. 


\vspace{0.2in}

We thank Sriram Ramaswamy and Andrea Cavagna for useful discussions. The research leading to this work was supported by the National Science Foundation (NSF) awards DMR-1305184  and DGE-1068780 at Syracuse University and NSF award PHY11-25915 and the Gordon and Betty Moore Foundation Grant No. 2919 at the KITP at the University of California, Santa Barbara. MCM also acknowledges support from the Simons Foundation. 

{\color{black} 

\section{Appendix A: Dimensionless parameters}

It is useful to clarify our choice of dimensionless parameters by making contact with the non-inertial Vicsek model familiar from the literature. The  continuous-time Vicsek model can be obtained from Eqs.~(1) and (2) of the main text by letting $\chi=0$ and eliminating $s_i$, with the result,
\begin{gather}
\label{eq:ri_a}
\frac{d\bm r_i}{dt}=v_0\bm \hat{\bm e}_{\theta_i};,\\
\label{eq:thetai_a}
\frac{d\theta_i}{dt}=\frac{\gamma}{\eta}\sum\limits_{j}\tilde{F}(\theta_j-\theta_i,\bm r_{ji})+\sqrt{\frac{2\epsilon}{\eta^2}}~ \xi_i(t)\;,
\end{gather}
where $\tilde{F}(\theta)=F(\theta)/(\pi R^2)$ for $|\mathbf{r}_{ij}|\leq R$ and zero otherwise, with $F(\theta)=\sin\theta$. The non-inertial limit corresponds to a Vicsek model with alignment strength $\gamma/\eta$ and noise amplitude $\epsilon/\eta^2$.   The additional parameters in Eqs.~\eqref{eq:ri_a} and \eqref{eq:thetai_a} are $v_0$, the mean density $\rho_0=N/L^2$, and the radius $R$ of the interaction. If we scale lengths with $R$ and times with $R/v_0$ the continuous time Vicsek model described by Eqs.~\eqref{eq:ri_a} and \eqref{eq:thetai_a}  contains three dimensionless parameters: the scaled noise, $\epsilon R/(\eta^2 v_0)$, the scaled alignment strength, $\gamma/(\eta Rv_0)$, and the mean density $\rho_0R^2$.   The discrete time Vicsek model can be recovered by  assuming that the alignment is instantaneous, i.e., $(\gamma/\eta R^2)^{-1}$ is short compared to all other times scales (specifically $\tau_\epsilon=\eta^2/\epsilon$ and the time step for updating the dynamics). The resulting model contain two dimensionless parameters: the mean density $\rho_0R^2$ and the noise $\epsilon R/(\eta^2 v_0)$, as expected.

Alternatively, in Eqs.~\eqref{eq:ri_a} and \eqref{eq:thetai_a} we can scale times with $\tau_\epsilon$ and lengths with $v_0\tau_\epsilon$. The microdynamics then takes the form
\begin{gather}
\label{eq:ri_a_scaled}
\frac{d\bm r_i}{dt}=\bm \hat{\bm e}_{\theta_i};,\\
\label{eq:thetai_a_scaled}
\frac{d\theta_i}{dt}=\frac{\gamma\eta}{\epsilon}\sum\limits_{j\in C_{\tilde{R}}}{F}(\theta_j-\theta_i)+\sqrt{2}~ \xi_i(t)\;,
\end{gather}
where $\mathbf{R}_i$ and time are now all dimensionless, $F(\theta)=\sin\theta$ and we have made explicit the dependence on the interaction range, with $C_{\tilde{R}}$ a circle of radius $\tilde{R}=R\tau_\epsilon/v_0$.
The mean field limit of these equations will only depend on the dimensionless parameter $\gamma\eta\rho_0/\epsilon=\tau_\epsilon/\tau_\gamma$.
There are, however, two additional parameters that provide cutoffs to the mean-field theory: the interaction range at small scales and the system size at large scales, both scale with $v_0\tau_\epsilon$. In other words, although  seemingly magically rewritten in terms of a single parameter, the model still contains three independent parameters. {\color{black} When comparing to solution of the nonlinear PDE's obtained in mean-field to the results of particle simulations where we set $R=1$ one should think of the density $\rho_0R^2$ and $\gamma\eta/(\epsilon R^2)$ as independent parameters.}

For the inertial continuous time model described by Eqs. (1) and (2) of the main text, the same transformation yields a mean field theory that contains only two dimensionless parameters, defined as $\tilde{\gamma}$ and $\tilde{\chi}$ in the main text. To these, however, we must add the two cutoffs at large and small scales. The non-inertial limit is recovered for $\tilde{\chi}=0$.

}

\section{Appendix B: Derivation of the hydrodynamic equations}
The Fokker-Planck equation for the one-particle probability density $P(\bm r,\theta,s,t)$ associated with Eqs. (1) and (2) of the main text is given by
\begin{widetext}
\begin{gather}
\dot{P}(\bm r,\theta,s,t)+\bm v_{\theta}\cdot\nabla P=-\frac{\partial}{\partial\theta}(\frac{1}{\chi}sP)+T(\theta,\bm r,t)\frac{\partial P}{\partial s}+\frac{\partial}{\partial s}(\frac{\eta}{\chi}sP)+\epsilon\frac{\partial^2 P}{\partial s^2}\:,
\end{gather}
\end{widetext}
where $T(\theta,\bm r,t)=-\gamma\int_{-\pi}^{\pi}d\theta' F(\theta'-\theta)p_0(\bm r,\theta',t)$ is the torque. We have assumed local interaction $F(\theta,\bm r)=\delta(\bm r)$sin$(\theta)$ and defined $p_0(\bm r,\theta',t)=\int_{s}P(\bm r,\theta',s,t)$~\footnote{This interaction does not describe the mean polarization deep in the ordered state. We have verified that better behaved models such as  $F(\theta)\sim\sin(\theta/2)$, give equations of the same structure and do not affect the qualitative behavior and instabilities.}. To make the notation more compact, we define the Fokker-Planck operator as
\begin{gather}
L_k=L_{rev}+L_{ir}\:,\\
L_{rev}=-\frac{s}{\chi}\frac{\partial}{\partial\theta}+T(\bm r,\theta,t)\frac{\partial}{\partial s}-\bm v_{\theta}\cdot\bm\nabla\:,\\
L_{ir}=\frac{\eta}{\chi}\frac{\partial}{\partial s}(s+s_{0}^2\frac{\partial}{\partial s})\:,
\end{gather}
where $L_{rev}$ and $L_{ir}$ represent the reversible and irriversible part of the Fokker-Planck operator respectively, and we have introduced the steady state value of the spin $s^2_{0}=\epsilon\chi/\eta$. In the absence of interaction and activity, the steady state distribution of the spin, obtained by setting the time derivative to zero, has a Maxwell-like form, given by
\begin{gather}
P_0(s)=\frac{1}{\sqrt{2\pi s^2_{0}}}\exp(-\frac{s^2}{2s^2_{0}}).
\end{gather}
Following standard methods~\cite{Risken1988}, we transform the Fokker-Planck operator by multiplying it from the right and the left by $\phi_0(s)=P_0^{\frac{1}{2}}(s)$ and $\phi_0^{-1}(s)=P_0^{-\frac{1}{2}}(s)$, respectively, with the result
\begin{gather}
\bar{L}_k=\phi_0^{-1}(s)L_k \phi_0(s)=\bar{L}_{rev}+\bar{L}_{ir}\:,\\
\bar{L}_{ir}=-\frac{\eta}{\chi} b^+b\:,~~\bar{L}_{rev}=-bD-b^+\hat{D}-\bm v_{\theta}\cdot\bm \nabla\:,
\end{gather}
where $b^{+}$ and $b$ are creation and annihilation operators, respectively. 
\begin{gather}
b^+\phi_n(s)=\sqrt{n+1}\phi_{n+1}(s)\:,\\
b\phi_n(s)=\sqrt{n}\phi_{n-1}(s).
\end{gather}
$D$ and $\hat{D}$ are the differential operators, with the latter containing the information of the interaction,
\begin{gather}
b^{+}=-s_{0}\frac{\partial}{\partial s}+\frac{1}{2}\frac{s}{s_{0}}\:,~~~~~b=s_{0}\frac{\partial}{\partial s}+\frac{1}{2}\frac{s}{s_{0}}\:,\\
D=\frac{s_{0}}{\chi}\frac{\partial}{\partial\theta}\:,~~~~~\hat{D}=\frac{s_{0}}{\chi}\frac{\partial}{\partial\theta}+\frac{T(\theta,\bm r,t)}{s_{0}}.
\end{gather}
The normalized eigenfunctions $\phi_n(s)$ of the operator $\bar{L}_{ir}=-\frac{\eta}{\chi} b^+b$ are defined by the eigenvalue equation
\begin{gather}
\bar{L}_{ir}\phi_n(s)=-\frac{\eta}{\chi} n\phi_n(s)\:,
\end{gather}
with
\begin{gather}
\phi_n(s)=(b^+)^n\phi_0(s)/\sqrt{n!}\:,\\
\phi_0(s)=\exp(-\frac{s^2}{4s^2_{0}})/\sqrt{s_{0}\sqrt{2\pi}}.
\end{gather}
Finally, $\phi_n(s)$ are related to the physicists' Hermite polynomials $H_n(x)=\left(2x-\frac{d}{dx}\right)^n\cdot 1$ as
\begin{gather}
\phi_n(s)=H_n(\frac{s}{\sqrt{2}s_{0}})\exp(-\frac{s^2}{4s^2_{0}})/\sqrt{n!2^n s_{0}\sqrt{2\pi}}.
\end{gather}
We now expand the probability distribution function in terms of $\phi_n(s)$,
\begin{gather}
P(\bm r,\theta,s,t)=\phi_0(s)\sum_{n=0}^{\infty}p_n(\bm r,\theta,t)\phi_n(s)\:,
\end{gather}
and we insert the expansion into the \black{Fokker-Planck} equation,
\begin{gather}
\partial_t P(\bm r,\theta,s,t)=L_k P(\bm r,\theta,s,t)\:,
\end{gather}
where the \black{Fokker-Planck} operator is obtained after an inverse transformation, as
\begin{gather}
L_k=\phi_0(s)(-\frac{\eta}{\chi} b^+b-bD-b^+\hat{D}-\bm v\cdot\bm \nabla)\phi_0^{-1}(s).
\end{gather}
Using the properties of the operators and the orthogonality of the Hermite polynomials, we obtain a hierachy of equations for the moments $p_n(\bm r,\theta,t)$,
\begin{widetext}
\begin{gather}
\label{eq:moments}
\mathcal{D}_t p_n(\bm r,\theta,t)=-\frac{\eta}{\chi} np_n(\bm r,\theta,t)-\sqrt{n+1}Dp_{n+1}(\bm r,\theta,t)-\sqrt{n}\hat{D}p_{n-1}(\bm r,\theta,t)\:,
\end{gather}
\end{widetext}
where $\mathcal{D}_t=\partial_t+\bm v_{\theta}\cdot\bm \nabla$ is the material derivative. Explicitly, the equations for the first three moments are given by
\begin{gather}
\mathcal{D}_t p_0=-Dp_1\:,\\
\mathcal{D}_t p_1=-\frac{\eta}{\chi} p_1-\sqrt{2}Dp_2-\hat{D}p_0\:,\\
\mathcal{D}_t p_2=-\frac{2\eta}{\chi} p_2-\sqrt{3}Dp_3-\sqrt{2}\hat{D}p_1.
\end{gather}
The first two moments are related to the probability density $c(\bm r,\theta,t)$ of finding a particle at $\bm r$, with velocity directed along $\theta$ at time $t$ and the spin current $j(\bm r,\theta,t)$ as
\begin{gather}
c(\bm r,\theta,t)=p_0=\int_{-\infty}^{\infty}P(\bm r,\theta,s,t)ds\:,\\
j(\bm r,\theta,t)=s_{0}p_1=\int_{-\infty}^{\infty}sP(\bm r,\theta,s,t)ds.
\end{gather}
To obtain closed equations for $c$ and $j$, we set $\mathcal{D}_t p_2=0$ for times long compared to $\chi/2\eta$, and let $p_n=0$ for $n\ge 3$. We then eliminate $p_2$ in favor of $p_0$ and $p_1$ to obtain closed equations. The equations for density and current are then given by
\begin{gather}
\mathcal{D}_t c(\bm r,\theta,t)=-\frac{1}{\chi}\frac{\partial j}{\partial \theta}\:,\\
\mathcal{D}_t j(\bm r,\theta,t)=-\frac{\eta}{\chi}j+\frac{\epsilon}{\eta^2}\frac{\partial^2 j}{\partial \theta^2}+\frac{1}{\eta}\frac{\partial [T(\bm r,\theta,t) j]}{\partial \theta}\\\notag-\frac{\epsilon}{\eta}\frac{\partial c}{\partial \theta}-T(\bm r,\theta,t) c.
\end{gather}
The goal is to obtain closed equations for the number density $\rho(\bm r,t)$, polarization density $\bm w(\bm r,t)$ and spin current $S(\bm r,t)$, which are the conserved, symmetry-breaking and relevant dynamic variables in the flocking system, respectively. Generalizing the method described in Ref.\cite{Bertin2009}, we introduce the angular Fourier transform of $c$ and $j$ as 
\begin{gather}
c_k(\bm r,t)=\int_{-\pi}^{\pi} c(\bm r,\theta,t)e^{ik\theta}d\theta\:,\\
j_k(\bm r,t)=\int_{-\pi}^{\pi}j(\bm r,\theta,t)e^{ik\theta}d\theta\:,
\end{gather}
which are related to $\rho(\bm r,t)$, $\bm w(\bm r,t)$ and $S(\bm r,t)$ by
\begin{gather}
\rho(\bm r,t)=c_0(\bm r,t)\:,~~~~S(\bm r,t)=j_0(\bm r,t)\:,\\
w_x(\bm r,t)=Re[c_1(\bm r,t)]\:,~~~~w_y(\bm r,t)=Im[c_1(\bm r,t)]\:,
\end{gather}
whose dynamic equations are
\begin{widetext}
\begin{gather}
\partial_t c_k(\bm r,t)+\frac{v_0}{2}\nabla^* c_{k+1}+\frac{v_0}{2}\nabla c_{k-1}=\frac{ik}{\chi}j_k\:,\\
\partial_t j_k(\bm r,t)+\frac{v_0}{2}\nabla^* j_{k+1}+\frac{v_0}{2}\nabla j_{k-1}=-\frac{\eta_k}{\chi}j_k+\frac{ik\epsilon}{\eta}c_k+\frac{ik\gamma}{2\pi\eta}\sum_{m}j_{k-m}F_{-m}c_{m}+\frac{\gamma}{2\pi}\sum_{m}c_{k-m}F_{-m}c_m\:,
\end{gather}
\end{widetext}
where $\nabla=\partial_x+i\partial_y$, $\nabla^*=\partial_x-i\partial_y$ and $F_{\pm 1}=\pm i\pi$. We have introduced an effective friction 
$\eta_k=\eta+k^2\epsilon\chi/\eta^2$.
Explicity, the equations for $c_0$, $c_1$ and $j_0$ are given by
\begin{gather}
\partial_t c_0+\frac{v_0}{2}\nabla^*c_1+\frac{v_0}{2}\nabla c^*_{1}=0\:,\\
\partial_t c_1+\frac{v_0}{2}\nabla^*c_2+\frac{v_0}{2}\nabla c_{0}=\frac{i}{\chi}j_1\:,\\
\partial_t j_0+\frac{v_0}{2}\nabla^*j_1+\frac{v_0}{2}\nabla j^*_{1}=-\frac{\eta}{\chi}j_0.
\end{gather}
To close these equations, we need to express $j_1$ and $c_2$ in terms of $c_0$, $c_1$ and $j_0$. To do so, we consider the equations for $j_1$, $j_2$ and $c_2$,
\begin{widetext}
\begin{gather}
\partial_t j_1+\frac{v_0}{2}\nabla^*j_2+\frac{v_0}{2}\nabla j_0=-\frac{\eta_1}{\chi}j_1+\frac{i\epsilon}{\eta}c_1-\frac{\gamma}{2\eta}(j_2c^*_1-j_0c_1)+\frac{i\gamma}{2}(c_2c^*_1-c_0c_1)\:,\\
\partial_t j_2+\frac{v_0}{2}\nabla^*j_3+\frac{v_0}{2}\nabla j_1=-\frac{\eta_2}{\chi}j_2+\frac{2i\epsilon}{\eta}c_2-\frac{\gamma}{\eta}(j_3c^*_1-j_1c_1)+\frac{i\gamma}{2}(c_3c^*_1-c_1c_1)\:,\\
\partial_t c_2+\frac{v_0}{2}\nabla^*c_3+\frac{v_0}{2}\nabla c_1=\frac{2i}{\chi}j_2.
\end{gather}
\end{widetext}
For times long compared to $\chi/\eta$, we set $\partial_t j_1=\partial_t j_2=0$. Retaining terms up to first order in $\chi/\eta$ we obtain the expression for $j_1$ and $j_2$,
\begin{widetext}
\begin{gather}
j_1=\frac{\chi}{\eta_1}\left[\frac{i\epsilon}{\eta}c_1+\frac{\gamma}{2\eta}j_0c_1+\frac{i\gamma}{2}(c_2c^*_1-c_0c_1)-\frac{v_0}{2}\nabla j_0\right]+\mathcal{O}(\chi^2),\\
\label{eqn:j2}
j_2=\frac{\chi}{\eta_2}(\frac{2i\epsilon}{\eta}c_2-\frac{i\gamma}{2}c_1^2)+\mathcal{O}(\chi^2).
\end{gather}
\end{widetext}
Inserting Eq.~\ref{eqn:j2} into the equation for $c_2$, we obtain,
\begin{gather}
\partial_t c_2+\frac{v_0}{2}\nabla^*c_3+\frac{v_0}{2}\nabla c_1=\frac{\gamma}{\eta_2} c_1^2-\frac{4\epsilon}{\eta\eta_2}c_2.
\end{gather}
For times long compared to $\eta\eta_2/(4\epsilon)$, we follow the method of Ref.~\cite{Bertin2009, Chate2014} and set $\partial_t c_2=0$ and $c_n=0$ for $n\ge 3$ to obtain the expression for $c_2$,
\begin{gather}
c_2=\frac{\gamma\eta}{4\epsilon}c^2_1-\frac{v_0\eta\eta_2}{8\epsilon}\nabla c_1.
\end{gather}
Using the expressions for $j_1$ and $c_2$, we obtain the closed equations,
\begin{widetext}
\begin{gather}
\frac{\partial c_0}{\partial t}+\frac{v_0}{2}\nabla c^*_1+\frac{v_0}{2}\nabla c_1=0\:,\\
\frac{\partial c_1}{\partial t}+\frac{v_0}{2}\nabla c_0+\frac{v_0\gamma\eta}{8\epsilon}\nabla^*c_1^2=\left(\frac{\gamma}{2\eta_1}c_0-\frac{\epsilon}{\eta\eta_1}-\frac{\gamma^2\eta}{8\epsilon\eta_1}|c_1|^2\right)c_1+\frac{i\gamma}{2\eta\eta_1}j_0c_1\notag\\-\frac{iv_0}{2\eta_1}\nabla j_0+\frac{\gamma v_0\eta\eta_2}{16\epsilon\eta_1}c^*_1\nabla c_1+\frac{v_0^2\eta\eta_2}{16\epsilon}\nabla^2 c_1\:,\\
\frac{\partial j_0}{\partial t}+\frac{i\chi\gamma v_0^2\eta\eta_2}{32\epsilon\eta_1}\left(\nabla\left[(\nabla^*c_1^*)c_1\right])-\nabla^*\left[(\nabla c_1)c_1^*\right]\right)=-\frac{v_0\gamma\chi}{4\eta_1}\left[i\nabla (c_0c^*_1)-i\nabla^*(c_0c_1)\right]-\frac{v_0\epsilon\chi}{2\eta\eta_1}\left(-i\nabla c^*_1+i\nabla^*c_1\right)\\-\frac{v_0\gamma^2\eta\chi}{16\epsilon\eta_1}\left[-i\nabla(|c_1|^2c^*_1)+i\nabla^*(|c_1|^2c_1)\right]-\frac{v_0\gamma\chi}{4\eta\eta_1}\left[\nabla (c^*_1j_0)+\nabla^* (c_1j_0)\right]+\frac{\chi v_0^2}{2\eta_1}\nabla^2 j_0-\frac{\eta}{\chi}j_0.
\end{gather}
\end{widetext}
Using the following identities,
\begin{gather}
\nabla^*c^2_1=\left[2(\bm w\cdot\nabla)\bm w+2\bm w(\nabla\cdot\bm w)-\nabla|w|^2\right]\:,\\
ic_1j_0=\bm S\times\bm w\:,\\
i\nabla j_0=-\nabla\times\bm S\\
i\nabla (c_0c^*_1)-i\nabla^*(c_0c_1)=-2\nabla\times(\rho\bm w)\:,\\
-i\nabla c^*_1+i\nabla^*c_1=2\nabla\times\bm w\:,\\
-i\nabla(|c_1|^2c^*_1)+i\nabla^*(|c_1|^2c_1)=2\nabla\times (|w|^2\bm w)\:,\\
\nabla (c^*_1j_0)+\nabla^* (c_1j_0)=2\bm S\nabla\cdot\bm w+2(\bm w\cdot\nabla)\bm S\:,\\
c^*_1\nabla c_1=(\bm w\cdot\nabla)\bm w-\bm w(\nabla\cdot\bm w)+\frac{1}{2}\nabla |\bm w|^2\:,\\
i\left(\nabla\left[(\nabla^*c_1^*)c_1\right])-\nabla^*\left[(\nabla c_1)c_1^*\right]\right)=-2\bm w\times\nabla^2\bm w,
\end{gather}
we finally obtain the hydrodynamic equations~\footnote{An alternative closure proposed in \cite{Bartolo2013} yields continuum equations with the same structure as those obtained here, but with different coefficients.},
\begin{widetext}
\begin{gather}
\frac{\partial \rho}{\partial t}=-\nabla\cdot(v_0\bm w)\:,\\
\frac{\partial \bm w}{\partial t}+\frac{v_0}{2}\nabla\rho+\frac{v_0\gamma\eta}{8\epsilon}\left[2(\bm w\cdot\nabla)\bm w+2\bm w(\nabla\cdot\bm w)-\nabla|w|^2\right]=
\left(\frac{\gamma}{2\eta_1}\rho-\frac{\epsilon}{\eta\eta_1}-\frac{\gamma^2\eta}{8\epsilon\eta_1}|w|^2\right)\bm w\\+\frac{\gamma}{2\eta\eta_1}\bm S\times\bm w+\frac{v_0}{2\eta_1}\nabla\times\bm S+\frac{\gamma v_0\eta\eta_2}{16\epsilon\eta_1}\left[(\bm w\cdot\nabla)\bm w-\bm w(\nabla\cdot\bm w)+\frac{1}{2}\nabla |w|^2\right]+\frac{v_0^2\eta\eta_2}{16\epsilon}\nabla^2\bm w\:,\\
\frac{\partial \bm S}{\partial t}=\frac{v_0\gamma\chi}{2\eta_1}\nabla\times(\rho\bm w)-\frac{v_0\chi\epsilon}{\eta\eta_1}\nabla\times\bm w-\frac{v_0\gamma^2\chi\eta}{8\epsilon\eta_1}\nabla\times(|w|^2\bm w)-\frac{v_0\gamma\chi}{2\eta\eta_1}\bm \left[\bm S(\nabla\cdot\bm w)+(\bm w\cdot\nabla) \bm S\right]\notag\\+\frac{\chi\gamma v_0^2\eta\eta_2}{16\epsilon\eta_1}\bm w\times\nabla^2\bm w+\frac{\chi v_0^2}{2\eta_1}\nabla^2\bm S-\frac{\eta}{\chi}\bm S.
\end{gather}
\end{widetext}
\section{Appendix C: Mode analysis}

We start with the dimensionless hydrodynamic equations. Time is scaled by the rotational diffusion time $\tau_{\epsilon}=\eta^2/\epsilon$ and length by the persistence length \black{$v_0\tau_{\epsilon}$}. $\bm w$ and $\rho$ are scaled by the average number density $\rho_0$ and $\bm S$ by $\rho_0\chi/\tau_{\epsilon}$, leading to~\footnote{If we neglect $\partial_t\bm S$ in Eq.~\eqref{eq:spin_dimless} and use the resulting equations to eliminate $\bm S$ in favor of $\rho$ and $\bm w$, the resulting continuum equations have the same structure as those obtained in \cite{Farrell2012}, with $\mathcal{O}(\tau_\eta/\tau_\epsilon)$ corrections to various coefficients.}
\begin{widetext}
\begin{gather}
\frac{\partial \rho}{\partial t}=-\nabla\cdot\bm w,\\
\frac{\partial \bm w}{\partial t}+\lambda_1(\bm w\cdot\bm \nabla)\bm w=-\left[\alpha(\rho)+\beta|w|^2\right]\bm w-\frac{1}{2}\bm \nabla\rho+\lambda_2\bm w(\bm \nabla\cdot\bm w)+\Omega_1\bm S\times\bm w+\Omega_2\bm \nabla\times\bm S+D_w\nabla^2\bm w,\\
\label{eq:spin_dimless}
\frac{\partial \bm S}{\partial t}+\lambda_s(\bm w\cdot\nabla)\bm S=-\bm \nabla\times\left[(\alpha(\rho)+\beta|w|^2)\bm w\right]+\Omega_3\bm w\times\nabla^2\bm w-\lambda_s\bm S(\bm \nabla\cdot\bm w)-\xi\bm S+D_s\nabla^2\bm S.
\end{gather}
\end{widetext}
All parameters are related to two microscopic dimensionless variables: the scaled alignment strength $\tilde{\gamma}=\tau_{\epsilon}/\tau_{\gamma}$ and inertia $\tilde{\chi}=\tau_{\eta}/\tau_{\epsilon}$, where $\tau_{\epsilon}=\eta^2/\epsilon$, $\tau_{\eta}=\chi/\eta$ and $\tau_{\gamma}=\eta/(\gamma\rho_0)$ are the three natural timescales in the system corresponding to rotational diffusion, frictional dissipation and alignment interaction. 

 We drop the tilde in the following discussion for simplicity of notation.
\begin{widetext}
\begin{gather}
\alpha(\rho)=\frac{1}{1+\chi}(1-\frac{\gamma\rho}{2}),~~\beta=\frac{1}{1+\chi}\frac{\gamma^2}{8},\notag\\
\Omega_1=\frac{\chi\gamma}{2(1+\chi)},~~\Omega_2=\frac{\chi}{2(1+\chi)},~~\Omega_3=\frac{\gamma}{16}(\frac{1+4\chi}{1+\chi}),\notag\\
\lambda_1=\frac{\gamma}{4}-\frac{\gamma}{16}(\frac{1+4\chi}{1+\chi}),~~\lambda_2=-\left[\frac{\gamma}{4}+\frac{\gamma}{16}(\frac{1+4\chi}{1+\chi})\right]\lambda_s=\frac{\chi\gamma}{2(1+\chi)},\notag\\
\xi=\frac{1}{\chi},~~D_w=\frac{1+4\chi}{16},~~D_s=\frac{\chi}{2(1+\chi)}.\notag
\end{gather}
\end{widetext}
To perform linear mode analysis, we restrict ourselves to the 2D planar case. The isotropic state is always linearly stable therefore trivial, and we focus on the uniformly polarized state for $\gamma>2$ with the direction of spontaneous broken symmetry along $\hat{x}$. Perturbing around the polarized state $\rho=1+\delta\rho$, $\bm w=w_0\hat{x}+\delta\bm w$ and $\bm S=\delta S\hat{z}$, we arrive at the linearized equations
\begin{widetext}
\begin{gather}
\frac{\partial \delta\rho}{\partial t}=-\nabla\cdot\delta \bm w,\\
\frac{\partial \delta\bm w}{\partial t}+\lambda_1 w_0\partial_x\delta\bm w=\left(\mu_1\delta\rho+\mu_2\delta w_x\right)w_0\hat{x}-\frac{1}{2}\bm\nabla\delta\rho+\lambda_2 w_0\hat{x}\bm\nabla\cdot\delta\bm w+\Omega_1\delta S\hat{z}\times w_0\hat{x}+\Omega_2\bm\nabla\times\delta S\hat{z}+D_w\nabla^2\delta\bm w,\\
\frac{\partial \delta S\hat{z}}{\partial t}+\lambda_sw_0\partial_x(\delta S_z\hat{z})=\nabla\times\left[(\mu_1\delta\rho+\mu_2\delta w_x)w_0\hat{x}\right]+\Omega_3 w_0\hat{x}\times\nabla^2\delta w-\xi\delta S\hat{z}+D_s\nabla^2(\delta S\hat{z}),
\label{eqns_linearized}
\end{gather}
\end{widetext}
where $\mu_1=-\partial_{\rho}\alpha=\frac{\gamma}{2(1+\chi)}$, $\mu_2=-2\beta w_0=-\frac{w_0\gamma^2}{4(1+\chi)}$ and $w_0=\left[(\frac{\gamma}{2}-1)\frac{8}{\gamma^2}\right]^{1/2}$.
\subsection{Longitudinal mode $q=q_x$}
Considering mode along the direction of broken symmetry, we obtain
\begin{widetext}
\begin{gather}
\sigma\delta\rho=-iq\delta w_x,\\
\sigma\delta w_x=\mu_1 w_0\delta \rho+\mu_2 w_0\delta w_x-\frac{iq}{2}\delta\rho+iq(\lambda_2-\lambda_1)w_0\delta w_x-D_w q^2\delta w_x,\\
\sigma\delta w_y=-iq\lambda_1 w_0\delta w_y+\Omega_1 w_0\delta S-iq\Omega_2\delta S- D_wq^2\delta w_y,\\
\sigma\delta S=-q^2\Omega_3 w_0\delta w_y-\xi\delta S-iqw_0\lambda_s\delta S-D_sq^2\delta S
\end{gather}
\end{widetext}
\paragraph{\bf{``Banding Instability"}} Notice that $\delta\rho$ and $\delta w_x$ decouple from $\delta w_y$ and $\delta S$, leading to the dispersion relation 
\begin{gather}
\sigma_l(q)=\frac{i\mu_1}{\mu_2}q+\frac{1}{\mu_2 w_0}\left[\frac{(\lambda_2-\lambda_1)w_0\mu_1}{\mu_2}+\frac{1}{2}-\frac{\mu_1^2}{\mu_2^2}\right]q^2+\mathcal{O}(q^3).
\end{gather}
Fluctuations in density and magnitude of polarization lead to the ``banding instability" close to the isotropic-polar phase transition as generally observed in polar active fluid, the condition of which is given by
\begin{gather}
\frac{(\lambda_2-\lambda_1)w_0\mu_1}{\mu_2}+\frac{1}{2}<\frac{\mu_1^2}{\mu_2^2}.
\end{gather}
In terms of the microscopic parameters, it reads
\begin{gather}
\gamma<\frac{8}{3}.
\end{gather}
\paragraph{\bf{Spin wave}} Dynamics of $\delta w_y$ and $\delta S$ gives rise to the spin wave, carrying the information of turning. Neglecting convections and diffusion, the dispersion relation for the spin wave is
\begin{gather}
\sigma_s^{\pm}=-\frac{\xi}{2}\pm c_sq\sqrt{\frac{[\xi/(2c_s)]^2}{q^2}-1},
\end{gather}
where
\begin{gather}
c_s=w_0\sqrt{\Omega_1\Omega_3}=\left[(\frac{\gamma}{2}-1)\frac{\chi(1+4\chi)}{4(1+\chi)^2}\right]^{\frac{1}{2}}
\end{gather}
is the wave speed.
\subsection{Transverse mode $q=q_y$}

\paragraph{\bf{Transverse instability}}Transverse mode is governed by the full coupled equations:
\begin{widetext}
$$\begin{pmatrix} 0&0&-iq&0\\\mu_1 w_0&\mu_2w_0-D_wq^2&iq\lambda_2 w_0&iq\Omega_2\\-\frac{iq}{2}&iq\lambda_sw_0&-D_wq^2&\Omega_1w_0\\-iqw_0\mu_1&-iqw_0\mu_2&-q^2\Omega_3w_0&-\xi-D_sq^2\end{pmatrix}\begin{pmatrix}\delta\rho\\\delta w_x\\\delta w_y\\\delta S\end{pmatrix}=\sigma_t(q)\begin{pmatrix}\delta\rho\\\delta w_x\\\delta w_y\\\delta S\end{pmatrix},$$
\end{widetext}
which leads to the dispersion relation once treated perturbatively in the long wavelength limit:
\begin{widetext}
\begin{gather}
\sigma_t^{\pm}(q)=\pm \frac{\sqrt{2}i}{2}q+\frac{1}{\xi}\left(\frac{w_0\mu_1\Omega_1}{2\mu_2}-\frac{\Omega_1\Omega_3 w_0^2}{2}-\frac{\lambda_2 w_0^2\Omega_1}{2}-\frac{D_w\xi}{2}\right)q^2+\mathcal{O}(q^3),
\end{gather}
\end{widetext}
\black{from which the condition for transverse instability is obtained as}
\begin{gather}
\label{eqn:transverse_cond}
\frac{w_0\Omega_1\mu_1}{\mu_2}-\lambda_2 w_0^2\Omega_1>D_w\xi+\Omega_1\Omega_3 w_0^2,
\end{gather}
or in terms of microscopic parameters
\begin{gather}
\gamma>\frac{(1+4\chi)(1+\chi)}{8\chi^2}+4.
\end{gather}
The phase diagram is plotted in Fig.2a in the main text, with quantitative agreement between the numerical and analytical phase boundaries. This transverse instability renders the system spatially inhomogeneous with large density and spin fluctuations characterized by continuously turning and swirling flocks with propagating spin waves. Therefore, we term it the spin-wave instability. The spatial-temporal patterns have been observed from both the numerical simulations of the hydrodynamic equations and particle simulations \footnote{We have performed extensive particle simulations that confirm the existence of a region of turning flocks and large spin density fluctuations at large $\gamma$. Typical snapshots from simulations are shown in Fig. 1 in the main text, but the full simulation results will be reported elsewhere.}. 

To understand the origin of the instability, we write down the minimal equations that yield this instability. For clarity, we write down the dimensionful form.
\begin{gather}
\label{eq:w}
\partial_t\bm w=-\beta|\bm w|^2\bm w+\lambda_2\bm w(\bm \nabla\cdot\bm w)+\Omega_1\bm S\times\bm w\;,\\ 
\partial_t\bm S=-v_0\chi\bm \nabla\times\left(\beta|w|^2\bm w\right)-\xi\bm S\;,
\label{eq:spin_simp_vect}
\end{gather}
where $\lambda_2=-v_0\gamma\eta/(4\epsilon)-\gamma v_0\eta\eta_2/(16\epsilon\eta_1)<0$, $\beta=\eta\gamma^2/(8\epsilon\eta_1)$, $\Omega_1=\gamma/(2\eta\eta_1)$ and $\xi=\eta/\chi$. The linearized equations are
\begin{gather}
\label{eq:w}
\partial_t\delta w_x=\mu_2 w_0\delta w_x+\lambda_2 w_0\partial_y\delta w_y\;,\\ 
\partial_t\delta w_y=\Omega_1 w_0\delta s_z\;,\\
\partial_t\delta s_z=-\mu_s w_0\partial_y\delta w_x-\xi\delta s_z\;,
\label{eq:spin_simp}
\end{gather}
where $\mu_2=-2\beta w_0<0$ and $\mu_s=v_0\chi \mu_2<0$. They lead to the dispersion relation
\begin{gather}
\label{eqn:transverse3}
\sigma_t(q)=-\frac{w_0^2\lambda_2\mu_s\Omega_1}{\xi\mu_2}q^2+\mathcal{O}(q^3),
\end{gather}
which yields the instability condition
\begin{gather}
\label{eqn:transverse3}
\frac{w_0^2\lambda_2\mu_s\Omega_1}{\xi\mu_2}<0.
\end{gather}
This condition can be interpreted as the growth of bend and splay deformations augmented by the spin wave. If we include the density-dependent alignment interaction, rotational diffusion and spin elasticity, all of which serve as stabalization factors, we recover the full condition \ref{eqn:transverse_cond}. The competition among these effects yields the spin-wave instability, which is model-dependent.
%
%

\bibliography{References_v2}
\end{document}